\documentstyle[preprint,aps,psfig]{revtex}
\tighten
\begin{document}
\draft
\preprint{\vbox{Submitted to Physical Review C}}
\title{Novel Methods for Determining Effective Interactions \\
        for the Nuclear Shell Model} 

\author{H. Mueller${}^{1}$, J. Piekarewicz${}^{2}$, 
        and J.R. Shepard${}^{1}$}
\address{${}^{1}$Department of Physics, 
         University of Colorado, Boulder, CO 80309}
\address{${}^{2}$Department of Physics, 
         Florida State University, 
         Tallahassee, FL 32306}
\date{\today}
\maketitle
 
\begin{abstract}
  The Contractor Renormalization (CORE) method is applied 
  in combination with modern effective-theory techniques 
  to the nuclear many-body problem. A one-dimensional---yet 
  ``realistic''---nucleon-nucleon potential is introduced 
  to test these novel ideas. It is found 
  that the 
  magnitude of  ``model-space'' (CORE) corrections 
  diminishes considerably when an effective potential that 
  eliminates the hard-momentum components of the potential 
  is first introduced. As a result, accurate predictions 
  for the ground-state energy of the there-body system are 
  made with relatively little computational effort 
  when both techniques are used in a complementary fashion.
\end{abstract}
\narrowtext

\section{Introduction}
\label{sec:intro}

The shell-model has long been the standard paradigm for treating the 
nuclear many-body problem. However, quite recently, the form of the 
shell model most frequently employed {\it in practice} has been 
criticized by Haxton and coworkers\cite{haxt99,haxt01}. The chief 
objection in these papers is that shell-model practitioners often 
use effective interactions which do not follow in any systematic 
way from realistic nucleon-nucleon (NN) interactions nor from the 
inevitable truncations that these entail. They point out that 
theoretical schemes exist that indicate how ``bare''interactions 
should be modified to account for effects of truncations. Moreover, 
these schemes emphasize the necessity of using consistently modified 
operators rather than the bare ones used in many shell-model studies.
Finally, they stress that tremendous recent advances in raw 
computational power and in numerical techniques now make it feasible 
to implement these theoretical schemes.

During the past few years, two of us (JP and JRS) have extensively
studied low-dimensional quantum spin systems, especially the so-called
spin-ladder materials~\cite{piek96,piek97,piek98a,piek98b,piek99}.  We
note that interest in the study of the ladder compounds has been
stimulated by suggestions that these deceptively simple materials
could exhibit some of the critical behavior believed to be responsible
for high-temperature superconductivity~\cite{dago96}.  In much of this
work, we have made use of a novel ``plaquette'' basis and of a
truncation scheme of the full Hilbert space that is suggested by the
structure of this basis. Central to this work has been the use of the
COntractor REnormalization (CORE) method of Morningstar and
Weinstein\cite{morn94}. We have relied on CORE for the construction of
an effective Hamiltonian and effective operators to be employed in
concert with the truncated basis. Our success in applying CORE to
important condensed-matter problems has inspired us to attempt
adopting the approach to the nuclear-shell model as it directly
addresses many of the concerns expressed in the preceding paragraph.

Another important development, the use of effective-field theories 
in nuclear physics, has been stimulated by novel ideas in modern
renormalization theory~\cite{lepa97,bean00}. Effective theories 
(ET's) are low-energy approximations to the exact (and, thus, 
very complicated) theory. One of the central tenets behind ET's 
is that a long-wavelength particle (of wavelength $\lambda$) 
should be insensitive to details of the potential at very short
($d\!\ll\!\lambda$) distances. Hence, it should be possible to 
replace the very complicated short-distance structure of the 
exact theory by a potential with a simpler short-range structure, 
while leaving the low-energy properties of the theory intact. 
The aim of this paper is to combine both approaches---CORE and 
effective theories---in a complimentary manner.

The paper has been organized as follows. In Sec.~\ref{sec:formal} we
outline the basic steps in the implementation of CORE to the nuclear
many-body problem. Particular emphasis is placed in the construction
of the similarity transformation that generates the effective
interaction.  In the same section the basic ideas behind modern
effective theories are introduced together with a detailed procedure
on how to merge these two (CORE and ET's) powerful approaches.  In
Sec.~\ref{sec:results} we illustrate how the effective potential is
constructed and the subsequent refinements of the basis that CORE
entails. In particular, we focus on predictions of the ground-state
energy of the three-body systems using a variety of approximations.
Finally, our conclusions and plans for future calculations are
presented in Sec.~\ref{sec:concl}.

\section{Formalism}
\label{sec:formal}

In this section we describe in detail how CORE is to be
used in the many-body problem. Central to the spirit of
CORE is a similarity transformation that ultimately
generates a renormalized Hamiltonian to be used 
in a severely truncated Hilbert space. We follow this 
discussion with a brief introduction to modern versions 
of effective theories placing particular emphasis on its 
use in combination with CORE.

\subsection{CORE for the Shell Model}
\label{sec:CORE}

Here we will briefly outline the CORE method as adapted for the 
nuclear-shell model. The underlying dynamics is described by a
non-relativistic Hamiltonian containing only two-body forces
\begin{equation}
  H_{A} = \sum_{n=1}^{A}\frac{{\bf p}_{n}^{2}}{2m} \,+
          \sum_{m<n=1}^{A} V({\bf r}_{m}\!-\!{\bf r}_{n}) \;,
 \label{HofA}
\end{equation}
where $V({\bf r}_{m}\!-\!{\bf r}_{n})$ is a realistic NN 
interaction. Because of potential problems with center-of-mass
motion, the above Hamiltonian is modified by the introduction 
of a harmonic term that depends exclusively on the coordinate 
of the center of mass ${\bf R}$: 
\begin{equation}
  V_{A}^{\Omega}({\bf R}) = 
  \frac{1}{2}mA\Omega^{2}{\bf R}^{2} \;.
 \label{VofR}
\end{equation}
A many-body basis is then constructed from products of single 
particle harmonic oscillator wave functions characterized by 
the same oscillator frequency $\Omega$. The modified $A$-body 
Hamiltonian, the standard starting point for most shell-model 
calculations\cite{navr00}, is now written as
\begin{equation}
  H_{A}^{\Omega} = \sum_{n=1}^{A}
  \left[
  \frac{{\bf p}_{n}^{2}}{2m}+\frac{1}{2}m\Omega^{2}{\bf r}_{n}^{2}
  \right] +
 \sum_{m<n=1}^{A}
  \left[
   V({\bf r}_{m}\!-\!{\bf r}_{n})-
  \frac{m\Omega^{2}}{2A}({\bf r}_{m}\!-\!{\bf r}_{n})^{2}
  \right] \;.
 \label{HofOmega}
\end{equation}
The terms depending on $\Omega$ allow separation of the center-of-mass 
motion and, after trivial subtractions of the center-of-mass energy, 
the calculations yield the desired physical spectrum which ultimately is
independent of $\Omega$. 

The first step in the implementation of CORE is the identification of
an ``elementary'' block and a truncation scheme~\cite{morn94}.  In our
case the elementary block, or one-body term, consists of the
single-nucleon Hamiltonian appearing in Eq.~(\ref{HofOmega}). This
one-body term, henceforth denoted by $h_1$, defines the basis via
products of one-body harmonic-oscillator eigenstates. Subsequently,
the truncation scheme is defined by retaining only those states with
energy less than some maximum (cutoff) value $E_{\rm max}$, or
equivalently oscillator quanta less than some maximum value $N$. In
this way it is guaranteed, by construction, that the spectrum of the
effective Hamiltonian $h_1$ reproduces identically the low-energy
properties of the exact one-body Hamiltonian.

The next step involves computation of the effective Hamiltonian for
the two-body system. To begin, the two-body problem must be solved
exactly. This is accomplished by using a very large basis, {\it i.e.},
a basis as large as can be afforded computationally and one
characterized by a cutoff $N_{\infty}$ which is effectively infinite.
For the 
method to have any utility we evidently require $N_{\infty}\gg N$. 
CORE now instructs us that the effective two-body Hamiltonian be given 
in terms of a similarity transformation: 
\begin{equation}
 H_{\rm eff}(1,2)\!\equiv\! S E_{\rm diag} S^{T}.
\end{equation}
where $E_{\rm diag}$ is the diagonal matrix containing the lowest 
${\cal N}$ exact eigenvalues of the two-body system where ${\cal N}$
is the number of two-body states defined by the cutoff $N$. 
Evidently, the structure of $H_{\rm eff}(1,2)$ guarantees,
independent of the details of the similarity transformation, 
the low-energy spectrum of the full theory will be preserved.
We now discuss the particular form of $S$ prescribed by CORE.

By elementary linear algebra, the columns of the similarity
transformation are the eigenvectors of $H_{\rm eff}(1,2)$.  CORE
demands that the first column of $H_{\rm eff}(1,2)$, {\it i.e.}, the
eigenvector associated with the lowest eigenvalue, has a non-zero
overlap with the lowest {\it exact} eigenvector of the full
Hamiltonian. A simple way to implement this demand is by choosing the
first eigenvector in the model space to be proportional to the lowest
exact eigenvector after it has been truncated and then properly
normalized. To construct the second column of the similarity
transformation one starts with the second exact eigenvector, truncates
it and then, via a Gram-Schmidt procedure, subtracts any component
parallel to the first column. This procedure guarantees the
fulfillment of another important CORE requirement: the $n^{\rm th}$
eigenvector in the model space must have a non-zero overlap with the
$n^{\rm th}$-lowest exact eigenvector yet no such overlap is allowed with
the previous $n-1$ eigenstates. The third column is
determined in a similar way and must have no component parallel to the
first two. This process is repeated until the ${\cal N}$ columns of
$S$ are completely determined. Note that, in practice, Gram-Schmidt is
numerically unstable and more sophisticated---but, in principle,
equivalent---techniques must be used\cite{morn94}. Finally the
effective two-body interaction is obtained by subtracting the one-body
terms from $H_{\rm eff}(1,2)$:
\begin{equation}
  V_{\rm eff}(1,2) =  H_{\rm eff}(1,2) - h_{1}(1) - h_{1}(2) \;.
 \label{veff2}
\end{equation}
Hence, the effective 2-body Hamiltonian to be used in the truncated 
model space is given by
\begin{equation}
  H_{\rm eff}^{(2)} = \sum_{n=1}^{A} h_{n} +
                      \sum_{m<n=1}^{A} V_{\rm eff}(m,n) \;.
 \label{heff2}
\end{equation}
It is possible to construct three-body, four-body, etc, interactions
by similar methods. Indeed, this is essential if many-body properties
are to be reproduced {\it exactly}. Note that
the induced many-body interactions appear as a consequence of 
truncations even when the original nuclear Hamiltonian contains at most
two-body terms as in Eq.~(\ref{HofA}). The usefulness of the approach
depends, of course, on being able to stop at a low order in the
cluster expansion while retaining acceptable accuracy.  

\subsection{CORE and Effective Theories}
\label{sec:coreeft}

Effective Theories (ET's) are now widely employed in many fields of
physics including nuclear theory\cite{bean00}. The philosophical
similarities between CORE and ET's are obvious: in both cases one
attempts to incorporate in a systematic fashion the contributions from
short-ranged (high momentum/energy) physics into effective
interactions and operators that are defined in a Hilbert space which
excludes the high momentum/energy states. Indeed, Bogner and
Kuo\cite{bogn00} have recently examined the connection between
CORE-like approaches, such as Lee-Suzuki~\cite{suzu80,suzu82,suzu83}
and Krenciglowa-Kuo~\cite{kren74} schemes on one hand and ET's on the
other. We are presently undertaking similar studies using concepts
from CORE and renormalization-group (RG) methods outlined by Birse 
and collaborators~\cite{birs97}. Our results will appear in a future 
publication.

In the present work we examine how CORE and the methods of ET can be
applied to the shell-model problem in a {\it complementary}
manner. The basic idea is as follows: the bare $NN$ interaction (see,
{\it e.g.}, Ref.~\cite{av18}) is pathological in the sense that it is
characterized by a very strong short-range repulsion. Obtaining
accurate many-body energies using the bare interaction therefore
requires huge bases. CORE and other similar methods are used to remedy
this problem.  We propose to solve this problem by first constructing
an effective $NN$ potential of a convenient form for use in
shell-model calculations. The main theoretical underpinning behind
effective theories is that a long-wavelength particle should be
insensitive to details of the short-range interaction. Thus, one adds
new (smoother) interactions to ``mimic'' the effects of the unknown
short-distance physics. Since no explicit short-range contributions
will appear in the effective potentials, one expects smaller bases to
be sufficient. Moreover, there is no reason that CORE cannot be used
together with an effective potential to further refine the basis, a
strategy that may result in substantially increased computational
efficiency.
  
        Our determination of the effective potential follows 
from the procedures outlined by Lepage~\cite{lepa97} and those 
later adapted by Steele and Furnstahl~\cite{ste98a,ste98b} to 
treat the $NN$ interaction. For simplicity we work in one 
dimension and treat all particles as as spinless and 
distinguishable; they are assumed to have identical masses of 
$m\!=\!939$~MeV. We assume a bare $NN$ interaction with the
same pathologies of a realistic interaction. That is, it is
given by the sum of a strong short-range repulsive exponential 
and a medium-range attractive exponential:
\begin{equation}
  V(x) =  V_{\rm s}\ e^{-m_{\rm s} |x|} 
       +  V_{\rm v}\ e^{-m_{\rm v} |x|} \; .
 \label{vbare}
\end{equation}
The two masses were chosen to be equal to $m_{\rm s}\!=\!400$~MeV and
$m_{\rm v}\!=\!783$~MeV, respectively. The strengths of the
potentials, $V_{\rm s}\!=\!-506$~MeV and $V_{\rm v}\!=\!+1142.49$~MeV,
were chosen to give a binding energy and point root-mean-square (rms)
radius for the symmetric (``deuteron'') state of 
$E_{\rm b}\!=\!-2.2245$~MeV and $r_{\rm rms}\!=\!1.875$~fm and
$\sqrt{\langle r^{2}\rangle}\!=\!1.875$~fm, respectively.  Note that
for the $NN$ system an odd-parity state bound by $-0.180358$~MeV is 
also found and that the three body ground state for this potential 
is bound by $-6.32$~MeV.

\begin{figure}[h]
\centerline{
 \psfig
{figure=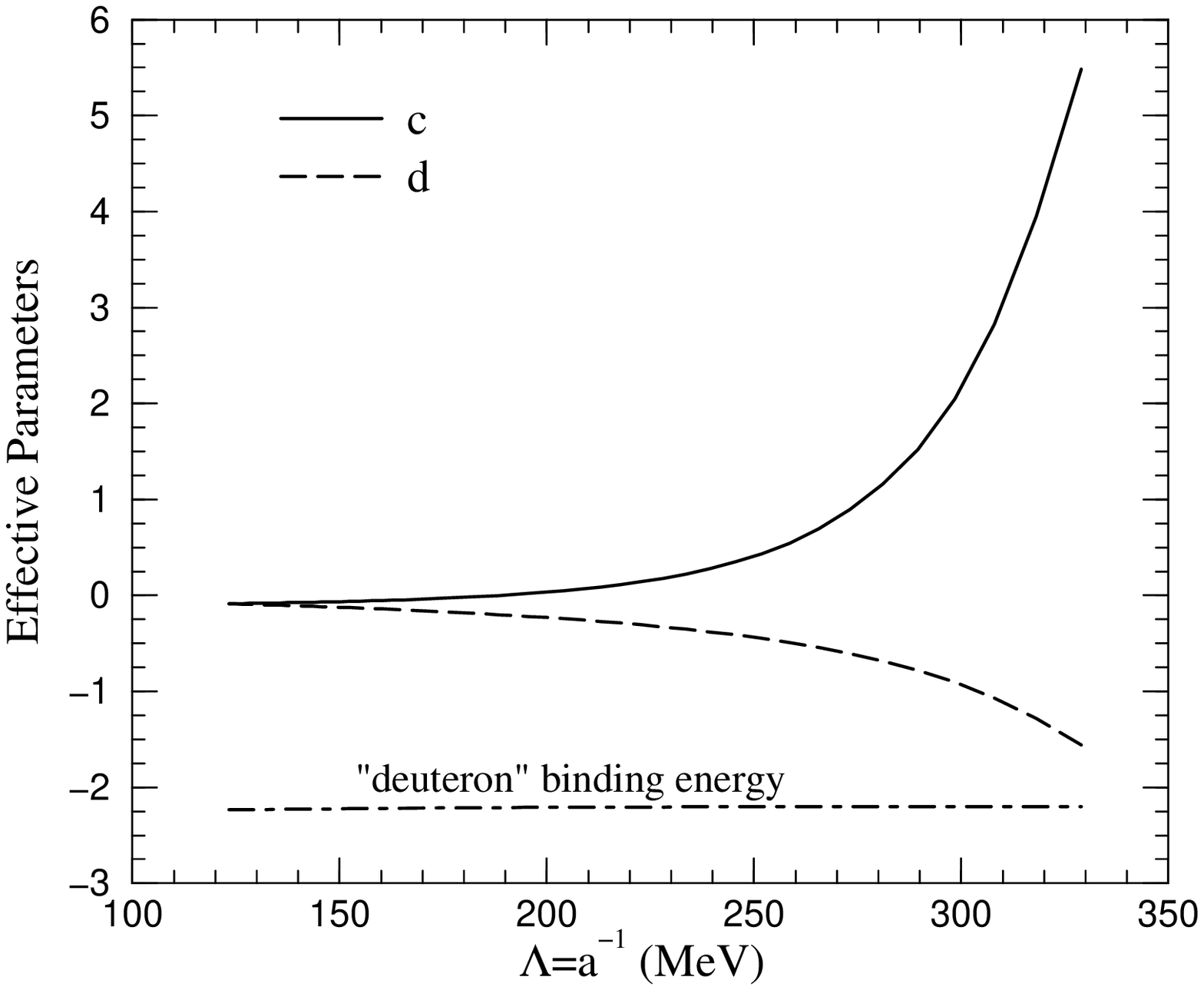,height=4in,width=4.5in,angle=00}}
\caption{The flow of the parameters of $V_{\rm eff}$ 
         as a function of the cutoff $\Lambda\!=\!1/a$.
         Also shown is the evolution of the two-body
         (deuteron) binding energy. The value of 
         $a=1.16$~fm (or $\Lambda=170$~MeV) was 
         chosen for all remaining calculations as 
         it reproduces the deuteron binding 
         energy exactly.}
\label{Fig1}
\end{figure}

Employing an option suggested by Lepage\cite{lepa97}, we choose a 
gaussian form for our effective potential:
\begin{equation}
  V_{\rm eff}(x) = \frac{1}{a}\left(c
                 + d\frac{\partial^2}{\partial \xi^2}
                 + e\frac{\partial^4}{\partial \xi^4}
                 + \ldots\right)\exp(-\xi^2)\;;
                   \quad \xi\equiv x/a\;.
 \label{veff}
\end{equation}
The range parameter has been fixed at $a$=1.16~fm as it reproduces the
deuteron binding energy exactly. Yet, as indicated in
Fig.~\ref{Fig1}, variations of this quantity---with the
corresponding adjustment of the effective parameters to reproduce the
low-energy phase shifts---have very little effect on the deuteron
binding energy and, indeed, on any of our calculations of low-energy
properties. Note that the parameters of the effective potential ($c$,
$d$, and $e$) are fixed separately for the even and odd channels. They
are constrained to reproduce the low-energy scattering phase shifts
for the appropriate channel, as we explained below.

\section{Results}
\label{sec:results}

\begin{figure}[h]
\centerline{
 \psfig
{figure=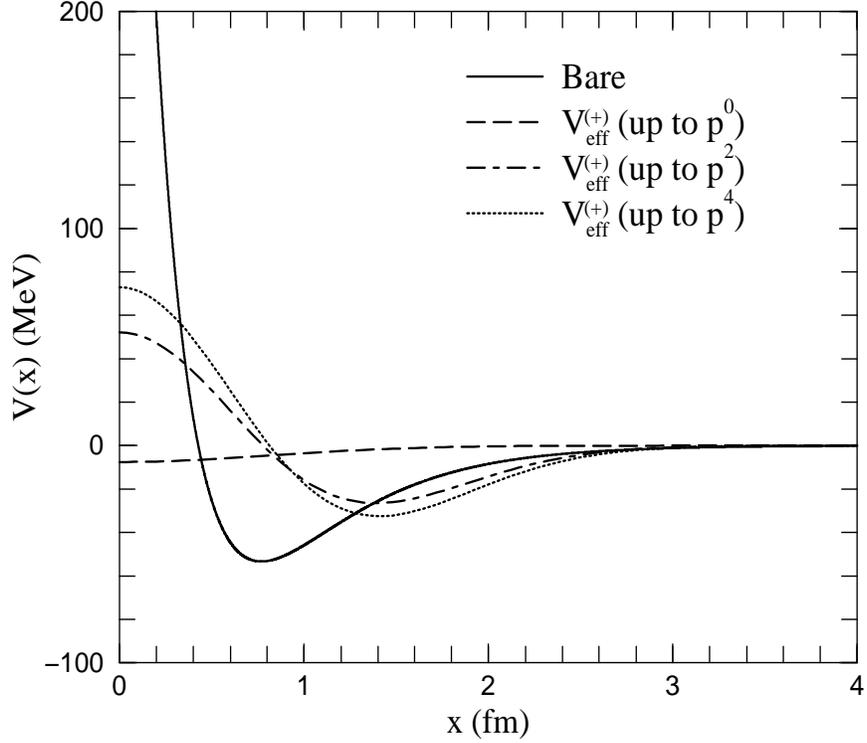,height=4in,width=4.5in,angle=00}}
\caption{The one-dimensional potentials used in our 
         two- and three-body calculations. The bare 
         potential yields a ``deuteron'' with the 
         correct binding energy and point radius 
         which qualitatively resembles the 
          $^1S_0$ $av18$-potential\protect\cite{av18}. 
          Also shown are the lowest (only $c\neq 0$ in
          Eqn.~\ref{veff}), next-to-lowest 
          (only $c$ and $d\neq 0$) and  
          next-to-next-to-lowest order (only $c$, 
          $d$ and $e\neq 0$) fits for the even channel.}
\label{Fig2}
\end{figure}

\begin{figure}[h]
\centerline{
 \psfig
{figure=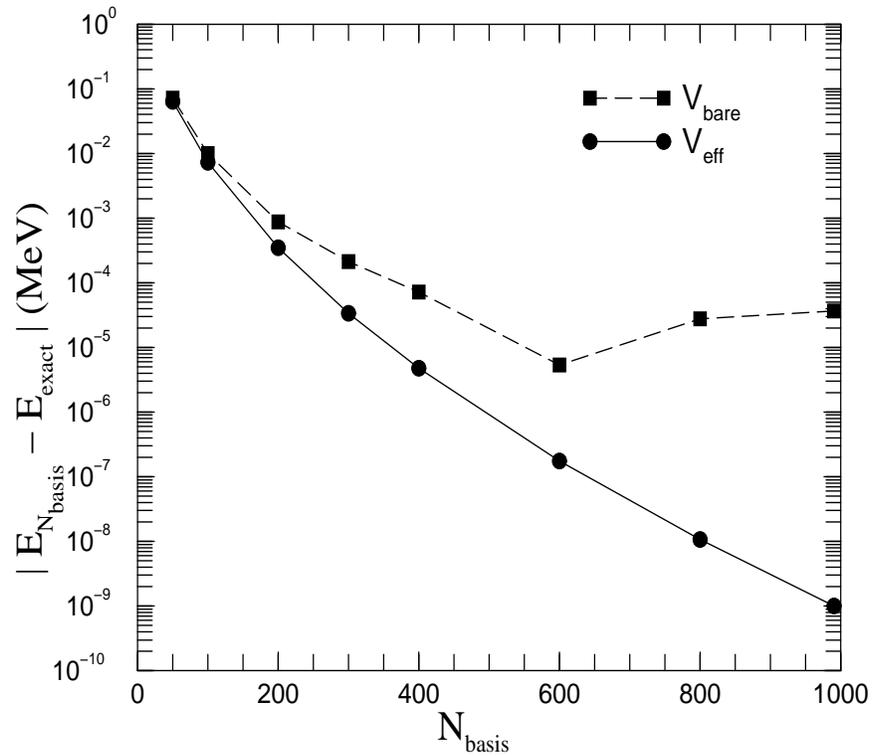,height=4in,width=4.5in,angle=00}}
\caption{Differences between the exact binding energy 
         of the one-dimensional deuteron as determined 
         by direct solution of the Schr\"odinger equation 
         and the energies determined by diagonalization in  
         harmonic-oscillator bases of various sizes are 
         displayed. Results using the bare and effective 
         potentials are compared and reveal superior 
         convergence using the latter.}
\label{Fig3}
\end{figure}

 Following the discussions in Ref.~\cite{ste98a,ste98b}, we fit the
parameters of $V_{\rm eff}$ 
to quantities which yield the effective 
range expansion when expanded in even powers of the wavenumber, $k$.
In one dimension we find for the even channel
\begin{equation}
        k\tan\delta_{\rm e}(k) = +\frac{1}{a_{\rm e}} 
                                 +\frac{r_{\rm e}}{2} k^2 
                                 +{\cal O}(k^{4}) \;.
 \label{even}
\end{equation}
where $\delta_{\rm e}$ is the even-channel phase shift and 
$a_{\rm e}$ and $r_{\rm e}$ are the scattering-length and 
effective-range parameters, respectively. This apparently 
unorthodox form of the effective-range expansion emerges 
from the fact that, in one dimension, the symmetric 
wave-function does not vanish at the origin. For the odd 
channel we find the familiar expression
\begin{equation}
        k\cot\delta_{\rm o}(k) = -\frac{1}{a_{\rm o}} 
                                 +\frac{r_{\rm o}}{2} k^2 
                                 +{\cal O}(k^{4}) \;.
 \label{odd}
\end{equation}
We have found that including only the the first two terms 
of Eq.~(\ref{veff}) for $V_{\rm eff}$ gives, as expected, 
an excellent reproduction of both scattering-lengths and 
effective-range parameters. The parameters of the effective 
potential are shown in Table~\ref{tableoneveff} along with 
the binding energies obtained using the bare and effective 
potentials. The resulting $V_{\rm eff}$'s for various orders 
of Eq.~(\ref{veff}) are compared with the bare potential in 
Fig.~\ref{Fig2}. Insofar as the low-energy properties 
of the theory are concern, the smooth $V_{\rm eff}$ is 
practically indistinguishable from the exact, and very 
pathological, $V_{\rm bare}$. As Table~\ref{tableoneveff} 
indicates, the effective potentials reproduce the exact 
binding energies very accurately. That the effective 
potentials do so without possessing the ``pathologies'' 
of the bare $NN$ potential alluded to earlier is 
demonstrated in Fig.~\ref{Fig3}. The exact binding 
energy for the even-parity state is determined accurately 
by solving the Schr\"odinger equation using the Numerov 
algorithm. This energy is also determined by evaluating 
the matrix elements of the potential in a harmonic-oscillator 
basis and then diagonalizing the matrix. Fig.~\ref{Fig3} 
shows that, as suggested above, such calculations using 
$V_{\rm eff}$ converge much faster than with the bare 
interaction where, eventually, round-off errors cause 
discrepancies to {\it grow} as the basis size increases.

\begin{figure}[h]
\centerline{
 \psfig
{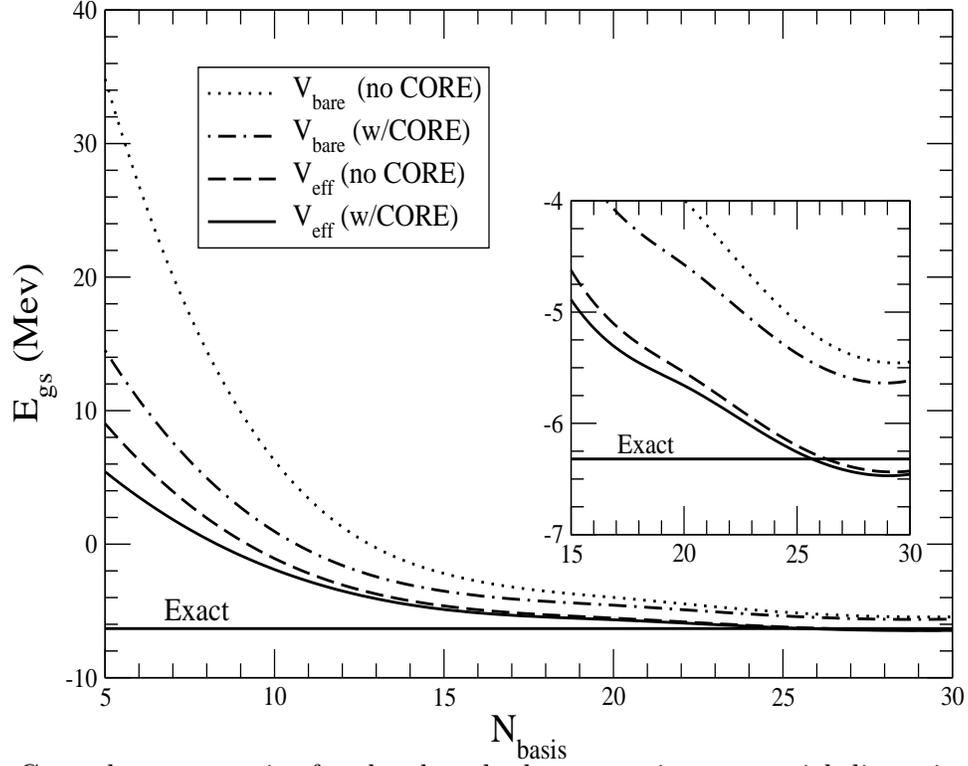}}
\caption{Ground-state energies for the three-body system 
         in one spatial dimension using the bare and effective 
         interactions with basis size determined by $N_{\rm basis}$. 
         Results are shown with and without using CORE. The exact 
         three-body ground-state energy is $-6.32$~MeV.}
\label{Fig4}
\end{figure}

Finally, we turn to the three-body system which represents the testing
ground for our model. Using the bare potential with very large basis
sizes ($N\simeq 100$) and then extrapolating to $N\rightarrow\infty$,
we find an exact ground-state energy for the three-body system of
$-6.32$~MeV.  Table~\ref{tablethreeveff} summarizes approximate
calculations of this quantity employing severely truncated bases; {\it
i.e.}, $N\leq 30$. (Note that the number of three-body states scales 
roughly like $N^3$.) The same information also appears in
Fig.~\ref{Fig4}.  Four types of calculations are reported:
calculations using either the bare or effective $NN$ interaction both
with and without CORE. The results show clearly that CORE corrections
are important, especially for small values of $N_{\rm basis}$. Yet
these corrections are significantly more important for the bare 
potential than for the effective one. This is to be expected as, in 
constructing the effective interaction, we have already ``softened'' 
the potential to such a degree that most of its large momentum 
components are absent. Put another way, the smoothed effective 
potential does not
explicitly contain the ``pathological'' short-range contributions
which CORE must treat. At the same time, the convergence to the exact
ground-state energy is much faster overall for the effective
interaction.  Note that this is true even when one considers that the
exact binding energy of the three-body system for the effective
potential ($-6.89$~MeV) differs slightly from the exact three-body
energy for the bare interaction ($-6.32$~MeV).  Presumably, this
difference is due to the omission of induced three-body 
interactions. This relatively small discrepancy suggests that
such three-body interactions are tractably small.

\section{Conclusions}
\label{sec:concl}

 A contractor renormalization (CORE) approach has been applied in
concert with modern versions of effective theories (ET's) to the
nuclear many-body problem. CORE is a ``model-space'' approach similar
to existing techniques, such as the Lee-Suzuki method, that have been
applied already with some success to the nuclear many-body
problem. The central concept behind CORE is a similarity
transformation responsible for generating a renormalized Hamiltonian
with identical low-energy properties to the exact Hamiltonian.  The
initial study presented here has been limited, for simplicity, to one
spatial dimension. Yet the bare nucleon-nucleon potential employed
throughout the paper possesses the same ``pathologies''---the
short-range repulsive core---as any realistic interaction. To remove
these pathologies we proceeded as follows.  First, the high-momentum
components of the potential were removed through the introduction of
an effective potential that was constrained -- in effect -- by 
low-energy scattering phase
shifts. Such an effective potential predicts a two-body (``deuteron'')
bound-state energy identical to that obtained with the bare
interaction. Second, in calculating the ground-state energy of the
three-body system we depended on CORE to compensate for severely 
truncating the size of
the basis employed in the calculation.  In computing the ground-state
energy of the three-body system we observed that, as in most
implementations of model-space techniques, CORE corrections
improve significantly the convergence towards the exact ground-state
energy.  However, use of the effective two-body potential significantly
accelerated the rate of convergence while simultaneously reducing the 
magnitude of the CORE corrections.

\begin{figure}[h]
\centerline{
 \psfig
{figure=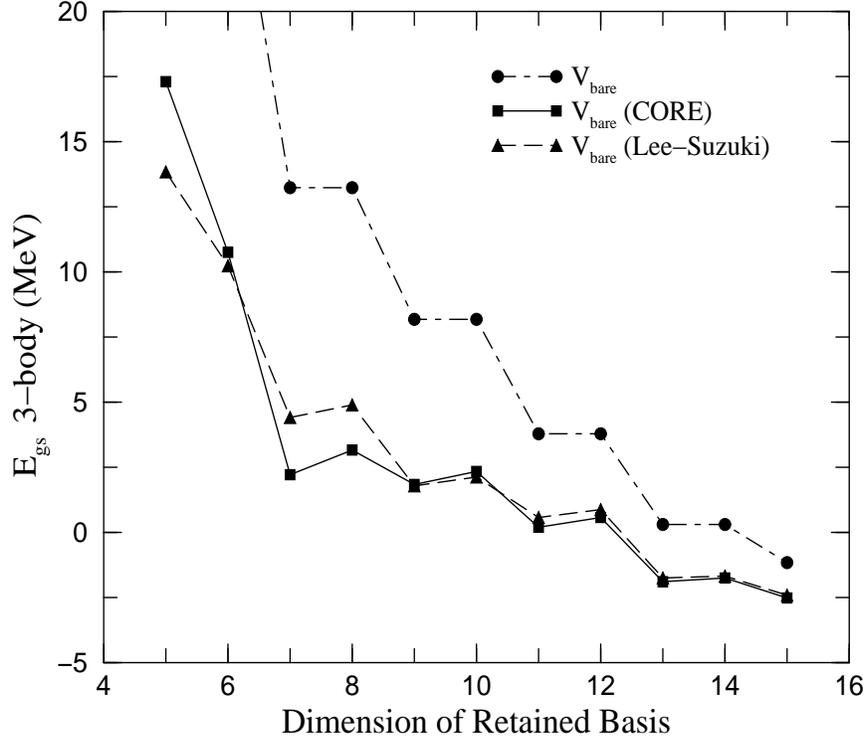,height=4in,width=4.5in,angle=00}}
\caption{Ground state energies for the 1D three-body system using
        relatively small basis sizes determined by 
        $N_{\rm trunc}$. Results using the uncorrected bare potential 
        are compared with those obtained using interactions corrected
        for truncations by two different model-space methods, namely
        CORE\protect\cite{morn94} and the more commonly used Lee-Suzuki 
        method\protect\cite{suzu82,suzu83}.
        All energies are in MeV; the exact three body ground state 
        energy is -6.32 MeV.}
\label{Fig5}
\end{figure}

Overall these results are gratifying as they suggest that ET's can be
profitably combined with more traditional model-space methods for use
with the nuclear shell model. For us, much remains to be done.  Apart
from the obvious extension to more realistic systems, we must
determine how to optimally mesh ET and CORE as well as how to correct
operators in a manner consistent with the construction of the
effective potential.  Also, a preliminary comparison between CORE and
a commonly used~\cite{navr00} approach for evaluating effective
interactions---the Lee-Suzuki method---has been made by us. We find 
that, at least for the one-dimensional three-body system treated here, 
the two methods give similar results, with ground-state energies 
computed using CORE being slightly---but consistently---closer to 
the exact value. (See Fig.\ref{Fig5} for relevant comparison with 
relatively small basis sizes.) In a future study we intend to elucidate the 
relationships between CORE, Lee-Suzuki, and other related approaches 
such as the Krenciglowa-Kuo scheme~\cite{kren74}. The formalism 
developed by Andreozzi~\cite{andr96} may prove useful in this regard.  
Finally, as mentioned earlier, it would be very useful and informative 
to find a set of RG-like equations\cite{birs97} which would lead 
directly to a sufficiently accurate version of $V_{\rm eff}$. 
Illuminating early steps in this direction may be found in
Refs.~\cite{haxt01,bogn00,bogn01a,bogn01b,schw01}.

\acknowledgments
We are grateful to the Institute for Nuclear Theory at
the University of Washington for their hospitality and 
support during the workshop on {\it Effective Field 
Theories and Effective Interactions.} It is also a 
pleasure to acknowledge very useful conversations with 
B. Barrett, W. Haxton, W. Leidemann, P. Navratil, and 
G. Orlandini. This work was supported by the DOE under 
Contracts Nos. DE-FC05-85ER250000, DE-FG05-92ER40750 
and DE-FG03-93ER40774. 

%
\newpage

\newpage

\newpage

\newpage

%
\mediumtext
 \begin{table}
   \begin{tabular}{ccccc}
     Channel  &  $c$  & $d$ & BE (bare) & BE (eff) \\
     \tableline
        Even & $-0.039342$ & $-0.159944$ 
             & $-2.2245$ MeV & $-2.2245$ MeV  \\
        Odd  & $-0.301863$ & $-0.104362$ 
             & $-0.1802$ MeV & $-0.1802$ MeV  \\
   \end{tabular}
  \vskip0.1in
  \caption{Parameters of $V_{\rm eff}$ [see Eq.~(\ref{veff})] 
           for even and odd channels. Also shown are the binding 
           energies using the bare and effective $NN$ interactions
           for a gaussian-cutoff value of $a\!=\!1.16$~fm.} 
  \label{tableoneveff}
 \end{table}

\vskip 0.5 true in

 \begin{table}
   \begin{tabular}{ccccc}
     $N_{\rm basis}$ & Bare--no CORE & Bare--with CORE & 
     Effective--no CORE & Effective--with CORE \\
     \tableline
      10 & $+6.2568$ & $+0.9258$ & $-1.0741$ & $-1.9028$  \\
      20 & $-3.9889$ & $-4.5934$ & $-5.5396$ & $-5.6624$  \\
      30 & $-5.4477$ & $-5.6327$ & $-6.4284$ & $-6.4606$  \\
   \end{tabular}
  \vskip .3in
  \caption{Ground-state energies for the one-dimensional three-body 
           system using the bare and effective interactions with 
           basis size determined by $N_{\rm basis}$. Results are 
           shown with and without using CORE. When CORE was used, 
           the size of the original basis was 
           $N_{\rm \infty}$=50. All energies are in MeV, with the 
           exact three-body ground-state energy being $-6.32$~MeV.}
  \label{tablethreeveff}
 \end{table} 

\end{document}